\documentclass[prl, 12pt]{revtex4-1}
\usepackage{amsmath}
\usepackage{amsfonts}
\usepackage{amssymb}
\usepackage{graphicx}
\usepackage[dvipsnames]{xcolor}

\usepackage{bm}
\usepackage{lmodern}
\usepackage{xspace}

\newcommand{\mol}[1]{\ensuremath{_{\text{#1}}}}
\newcommand{\cmol}[2]{\ensuremath{_{\text{#1}}^{\text{#2}}}}
\newcommand{\mr}[1]{\ensuremath{\mathrm{#1}}}

\newcommand{\tss}[1]{\ensuremath{^{\text{#1}}}}

\newcommand{\rrscan}{r\tss{2}SCAN\xspace}
\newcommand{\rrscanc}{r\tss{2}SCAN\ensuremath{_\mr{c}}\xspace}

\begin{document}

\title{A paradigm system for strong correlation and charge transfer competition}

\author{James W. Furness}
\email{jfurness@tulane.edu}
\author{Ruiqi Zhang}
\author{Jianwei Sun$^*$}
\email{jsun@tulane.edu}
\affiliation{Department of Physics and Engineering Physics, Tulane University, New Orleans, LA 70118, USA}

\date{\today}

\begin{abstract}
In chemistry and condensed matter physics the solution of simple paradigm systems, such as the hydrogen atom and the uniform electron gas, plays a critical role in understanding electron behaviors and developing electronic structure methods. The H\mol{2} molecule is a paradigm system for strong correlation with a spin-singlet ground state that localizes the two electrons onto opposite protons at dissociation. We extend H\mol{2} to a new paradigm system by using fractional nuclear charges to break the left-right nuclear symmetry, thereby enabling the competition between strong correlation and charge transfer that drives the exotic properties of many materials. This modification lays a foundation for improving practical electronic structure theories and provides an extendable playground for analyzing how the competition appears and evolves.
\end{abstract}


\maketitle

Once the governing quantum mechanical equations for a system of electrons under external potentials have been written down the electronic structure of the system is, in principle, determined. In practice however, almost all electron systems must be approximated to some degree as analytical solutions seldom exist when more than one electron is present. The paradigm systems with exact (or near exact) solutions are critical for understanding electron behaviors, formulating concepts, and developing electronic structure methods. For example, the exact electronic solutions of H and H\cmol{2}{+} are the foundations for quantum chemistry illustrating the concepts of atomic and molecular orbitals respectively. The uniform electron gas is a playground for condensed matter and many-body physics, whose accurate solution using Quantum Monte Carlo (QMC) methods \cite{Ceperley1980} enabled construction of the local spin density approximation (LSDA) \cite{Vosko1980, Perdew1992a, Sun2010} that supports most development in density functional theory (DFT) \cite{Hohenberg1964, Kohn1965}.

The H\mol{2} molecule is a paradigm system with a spin-singlet ground state in which the two electrons become strongly correlated as the bond is stretched to dissociation \cite{Ashcroft1976, Perdew2021a}. This is directly analogous to the strong correlations that emerge when different nuclei compete for valence electrons, particularly in transition metal materials. In transition metal oxides for example, adjacent metallic sites can compete for the valence $d$ electrons causing $d_i^nd_j^n \to d_i^{n-1}d_j^{n+1}$ charge fluctuations, with $i$ and $j$ labeling separate metallic sites. Such charge fluctuation involves the $d$-$d$ on-site Coulomb interaction $U$ that characterizes strong correlation between the $d$ electrons \cite{Hubbard1963}. Transition metal oxides host another type of charge fluctuation resulting from competition between the oxygen and transition metal ions for the valence electrons: $d_i^n \to d_i^{n+1}\underbar{O}$, where $\underbar{O}$ is a hole in the oxygen valence band. This type of fluctuation characterizes the charge transfer $\Delta$ between the metal and the anion valence band. Using $U$ and $\Delta$ as input parameters, Zaanan, Sawatzky, and Allen have been able to calculate a metal-insulator-transition phase diagram (ZSA) for transition metal compounds \cite{Zaanen1985}, though direct calculation of $U$ and $\Delta$ is challenging for most electronic structure theories. Furthermore, a recent QMC study has shown the pure 2D Hubbard model that only includes $U$, a celebrated effective Hamiltonian model for studying high critical temperature ($T_c$) cuprate superconductors, is insufficient for capturing superconductivity in the physically important parameter regime \cite{Qin2020a}.

While H\mol{2} presents a useful paradigm system for strong correlation, the symmetric nuclear potential from the equivalent hydrogen nuclei suppresses charge transfer, as both nuclei exert an equal pull on the two electrons. This left-right symmetry can be broken by replacing one proton with a helium nucleus, thus enabling charge transfer, driving both electrons to localize on the heavier He nucleus, and suppressing the strong correlation. This full replacement of strong correlation by charge transfer is unfortunate, since the competition between them drives exotic properties in many materials. It would be desirable to have a 2-electron paradigm system that can facilitate this competition, but the above analysis indicates that no such system can exist in nature given that all nuclei have integer charges. Here, we fill this role by extending H\mol{2} to a new paradigm system in which the competition between strong correlation and charge transfer can be tuned continuously.

To engineer such a system we create an asymmetric nuclear potential by replacing the hydrogen nuclei with fictitious fractional nuclear charges, $Z_A$ and $Z_B$. This enables charge transfer as $Z_A/Z_B$ moves away from 1 without completely suppressing strong correlation. Much like the uniform electron gas that fertilized enormous concepts in condensed matter physics, it is of no importance that the proposed system is fictional, since the physics it captures remain deeply relevant to real correlated materials. To make the two-electron system charge neutral we choose $Z_A + Z_B = 2$ and, without loss of generality, we require $Z_A \leq Z_B$ for simplicity. We term this as ``H\cmol{2}{FNC}'', where FNC stands for ``fractional nuclear charge''. Related systems were investigated by Cohen and Mori-S\'anchez in the context of the DFT derivative discontinuity and delocalization error in Ref \citenum{Cohen2014a}.

For the H\cmol{2}{FNC} paradigm system to be of any use it requires an exact solution. Hartree--Fock (HF) theory is exact for single electron systems such as H and H\cmol{2}{+} (up to the chosen basis set), but it is insufficient for systems of multiple electrons. It is difficult in general to obtain exact solutions for multi-electron systems, normally requiring exponentially scaling methods such as full configuration-interaction (FCI) or QMC. Fortunately, the two electron systems of interest here are small enough such that coupled-cluster at the singles-doubles (CCSD) level is equivalent to FCI, considering all possible excitations. We note however, that CCSD is not generally reliable for strongly correlated systems with more than two electrons. It is precisely this easy availability of exact solutions that make paradigm systems valuable assets for accessing the underlying physics of complex problems.

\begin{figure}
    \centering
    \includegraphics[width=\textwidth]{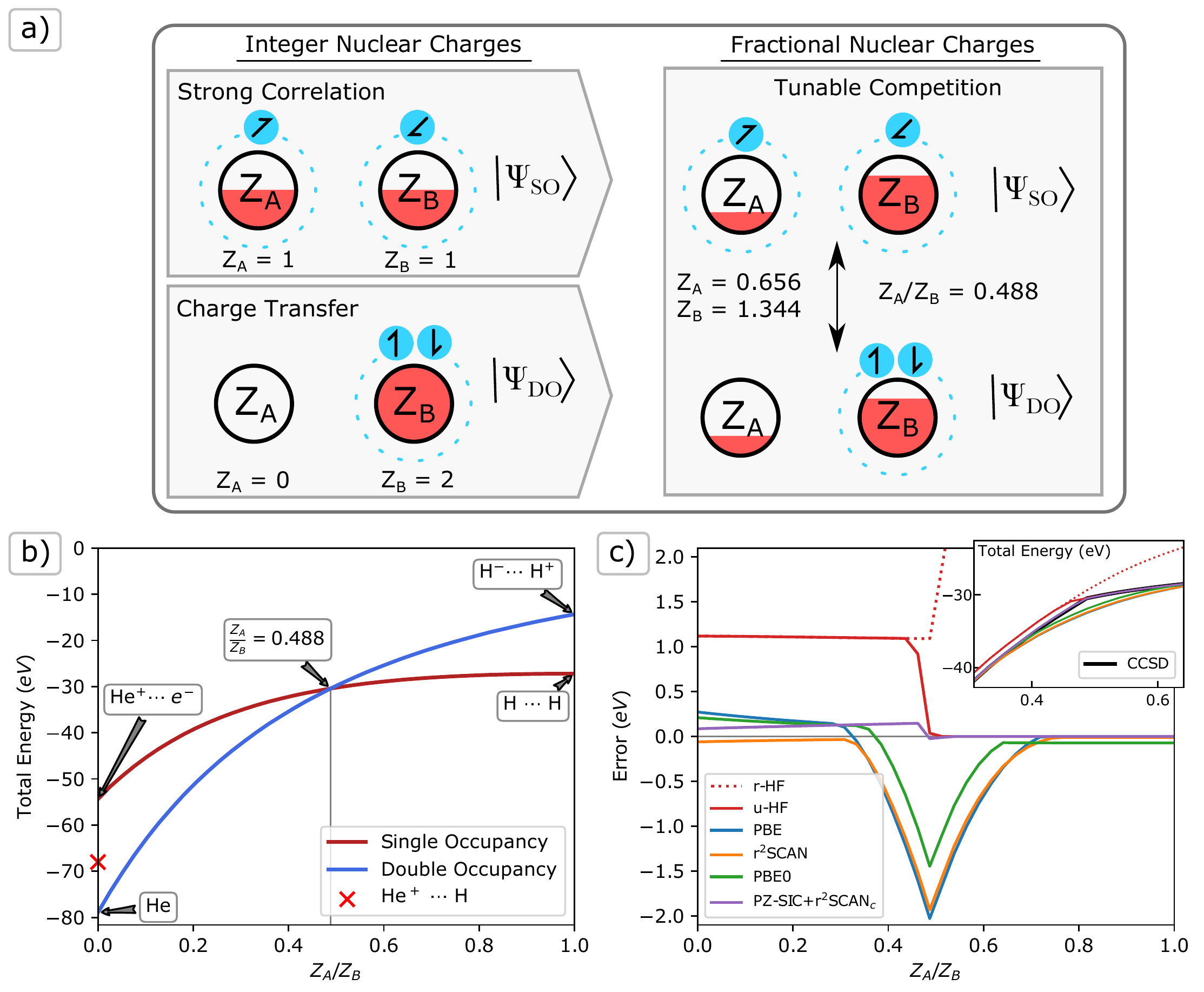}
    \caption{\textbf{Competition between charge transfer and strong correlation illustrated by H\cmol{2}{FNC} at infinite separation.} (a) A schematic plot of the single occupancy (SO) $\left|\Psi_\mathrm{SO}\right \rangle$, and double occupancy (DO) $\left|\Psi_\mathrm{DO}\right \rangle$ configurations in the spin-singlet state for integer and fractional nuclear charges. The nuclear charges $Z_A$ and $Z_B$ are constrained such that $Z_A + Z_B = 2$ with fractional charge represented by the red area within a circle of area 2. The blue discs represent electrons with spins labeled as arrows, where tilted arrows denote spin up-down degeneracy. (b) Exact total energies for SO and DO configurations as a function of the nuclear charge ratio, $Z_A/Z_B$. Within annotations, ``$\cdots$'' denotes infinite separation. (c) The errors in ground state energy calculated by HF and DFT approximations. Spin symmetry breaking (u-, solid lines) and conserving (r-, dotted lines) ground states are given for HF, while all DFT calculations are spin symmetry breaking. The corresponding total ground state energies are presented in the inset.}
    \label{fig:frac_atoms}
\end{figure}

There are two possible electronic configurations for the H\cmol{2}{FNC} system when the fractional nuclei are infinitely separated: a single occupation (SO) solution, $\left|\Psi_\mathrm{SO}\right \rangle$, with one electron on each nucleus characterizing strong correlation, and a double occupation (DO) solution, $\left|\Psi_\mathrm{DO}\right \rangle$, with both electrons on the more charged $Z_B$ nucleus characterizing charge transfer. Depending on the ratio $Z_A/Z_B$, the ground state is either the $\left|\Psi_\mathrm{SO}\right \rangle$ or $\left|\Psi_\mathrm{DO}\right \rangle$, or the configurations can become degenerate. This is summarized in Figure \ref{fig:frac_atoms} a), while Figure \ref{fig:frac_atoms} b) shows how the energy of the two configurations changes across the range of $0 \leq Z_A/Z_B \leq 1$, resulting in a discontinuous ground state. At $Z_A/Z_B = 1$, the ground state $\left|\Psi_\mathrm{SO}\right \rangle$ is comprised of separated neutral hydrogen atoms, while $\left|\Psi_\mathrm{DO}\right \rangle$, a H\tss{-} ion and a proton, is higher in energy by about 12 eV. At $Z_A/Z_B = 0$, $Z_B$ becomes a Helium nucleus and $Z_A$ disappears so the converse is true: $\left|\Psi_\mathrm{DO}\right \rangle$, a lone neutral helium atom, is favored over $\left|\Psi_\mathrm{SO}\right \rangle$, a He\tss{+} ion and a free electron. This is similar to the dissociation behavior of HHe\tss{+}. Allowing fractional nuclear charges forms a continuum between these limits with the energy of each configuration varying smoothly as the strong correlation of $\left|\Psi_\mathrm{SO}\right \rangle$ competes with the charge transfer of $\left|\Psi_\mathrm{DO}\right \rangle$. The $\left|\Psi_\mathrm{SO}\right \rangle$ and $\left|\Psi_\mathrm{DO}\right \rangle$ configurations become degenerate at $Z_A/Z_B \approx 0.488$, at which point the strong correlation and charge transfer competition is maximized. 

The appearance of degeneracy between $\left|\Psi_\mathrm{SO}\right \rangle$ and $\left|\Psi_\mathrm{DO}\right \rangle$ has profound implications. It is well known that neither the 1D Hubbard model nor its material realization in the infinite 1D hydrogen chain shows charge transfer physics \cite{Hachmann2006, Stella2011, Motta2017}. At short inter-atomic distances the hydrogen chain is weakly correlated and metallic, while at larger inter-atomic distances it undergoes a phase transition to a strongly correlated insulating phase  \cite{Hachmann2006, Stella2011, Motta2017}. The hydrogen chain is therefore a prototypical system illustrating the Mott-Hubbard metal insulator transition. Now, consider a hydrogen chain that has a sufficiently large inter-atomic distance such that the electron density overlap between atomic sites is negligible. Analogous to H\cmol{2}{FNC}, we can allow the nuclear charge for a pair of hydrogen atoms in the chain to be fractional under the constraint the whole system remains charge neutral. Then, following the previous analysis, the fractional nuclear charges can be tuned to make the pair close to the SO and DO degeneracy discussed above. Around this degeneracy a small perturbation, e.g., an electric field that enhances the potential at the more positive nucleus of the pair, can easily drive the electron from the less positive to the more positive nuclear site. This charge transfer capability under small perturbation emerging from the insulating hydrogen chain highlights the rich physics brought by the fractional nuclear charge. If more fractionally charged pairs are present then more degenerate states can be generated by tuning the fractional nuclear charges, potentially leading to exotic properties including superconductivity, even within this 1D model \cite{Fradkin2015}.

The exact solutions established above can highlight important deficiencies in common approximate electronic structure techniques. Here we focus on DFT, which has become a mainstay of computational materials studies. In principle, DFT is exact for the ground state energy and electron density through an efficient mapping of the interacting-electron problem onto an auxiliary non-interacting electron system described by a single determinant. In practice however, DFT methods must approximate the exchange-correlation energy functional that carries the many-electron effects. Paradigm systems have played critical roles in the development of exchange-correlation approximations \cite{Perdew1992a, Perdew1996, Sun2015, Furness2020a}, with each greatly enhancing the functional's predictive power when smoothly incorporated with other constraints. This role is played by the uniform electron gas for LSDA \cite{Vosko1980, Perdew1992a, Sun2010} and the hydrogen atom for the strongly-constrained and appropriately-normed  (SCAN) density functional and its \rrscan revision \cite{Sun2015, Furness2020a}. Hartree--Fock (HF) theory is also included here for comparison as it uses a single determinant to directly approximate the correlated wave function and is a base upon which many more sophisticated methods are built.

The exact ground state wave function of the H\cmol{2}{FNC} system must be a spin singlet configuration that maintains spatial symmetry between the spin-up and spin-down electrons. For single-determinant based methods however, it is energetically advantageous to allow different spin electrons to have different spatial distributions. This breaks the spin symmetry and handles the strong correlation of the ground state wave function by suppressing the spin fluctuation. The energetic benefit of using spin symmetry breaking can be seen when the ground state adopts the strongly correlated SO configuration, as shown in Figure \ref{fig:frac_atoms} (c) through the comparative errors of spin symmetry conserving ``restricted'' HF (r-HF) and spin symmetry breaking ``unrestricted'' HF (u-HF). The error of the r-HF ground state becomes large when strong correlation dominates in $Z_A/Z_B \gg 0.488$ where u-HF gives essentially exact ground state energies by localizing the spin-up and spin-down electrons onto separate nuclei. Localizing the electrons makes the overall system an effective sum of two one-electron sub-systems that are well described by a single determinant. The symmetry broken u-HF wave function is no longer an eigenfunction of the $\hat s^2$ spin operator however, and erroneously results in non-zero spin densities. The energetic benefit of symmetry breaking is therefore obtained at the expense of incorrect spin densities and a spin contamination in the single determinant solution. These pathologies are well known in studies of H\mol{2} \cite{Gunnarsson1976}. Recently, the symmetry broken solutions revealing the strong correlation have been interpreted as ``freezing'' a fluctuation in the exact correlated ground state wave function \cite{Perdew2021a}. Given this interpretation and the improved energies for strongly correlated systems we adopt the spin symmetry breaking strategy for DFT approximations to be discussed below.

We have selected four non-empirical DFT exchange-correlation functionals as examples from different levels of the Perdew--Schmidt hierarchy \cite{Perdew2001}. The Perdew--Burke--Ernzerhof (PBE) functional \cite{Perdew1996} is a standard at the generalized gradient approximation (GGA) level and is the simplest semi-local functional featured, taking only the spin density and its gradient as inputs. The meta-GGA level (the most sophisticated semi-local level) is represented by our recent \rrscan functional \cite{Sun2015, Furness2020a}, which includes the non-negative kinetic energy density as an additional ingredient that can be used to satisfy more exact constraints. Beyond the semi-local functionals we take the PBE0 functional \cite{Perdew1996, Adamo1999}, which replaces $25\%$ of the PBE exchange with $25\%$ of the non-local exact exchange of HF. Finally, we include the \rrscan functional with the Perdew--Zunger self-interaction correction (PZ-SIC+\rrscanc) \cite{Perdew1981} in which the self-interaction error is removed on an orbital-by-orbital basis, equivalent to HF+\rrscanc in a single orbital system such as H\cmol{2}{FNC}. The self-interaction error is a result of an incomplete cancellation of the self-Coulomb-repulsion by the self-exchange-interaction of orbitals in DFT approximations. When PZ-SIC is combined with a one-electron self-correlation free functional, such as \rrscanc, then the resulting DFT calculation is exact for one-electron systems such as H\mol{2}\tss{+}.

Figure \ref{fig:frac_atoms} (c) shows that DFT approximations are accurate for both the DO-dominant (small $Z_A/Z_B$) and SO-dominant (large $Z_A/Z_B$) states, for the latter of which the spin symmetry breaking is important. When $Z_A/Z_B \ll 0.488$, $\left|\Psi_\mathrm{DO}\right \rangle$ localizes both electrons onto the more charged $Z_B$ and u-HF matches r-HF yielding a total energy $\sim$1.1 eV too high as the short-range dynamic correlation is missed. In contrast, DFT approximations capture the short range dynamic correlation, delivering total energies within 0.25 eV of the reference. Severe errors are found for the DFT approximations without self-interaction correction when $Z_A/Z_B$ is in the region closely surrounding the degenerate point, around $0.3 < Z_A/Z_B < 0.7$. Here, the self-interaction error results in a spurious charge delocalization with one electron becoming shared across both nuclei. This leads to a ground state configuration with incorrectly fractional electron occupation numbers, even when the two nuclei are infinitely separated \cite{Perdew1982, Mori-Sanchez2008, Cohen2014a}. As a result there is an erroneously smooth connection between the $Z_A/Z_B \to 0$ and $Z_A/Z_B \to 1$ limits, with no discontinuity in the ground state energy at the $Z_A/Z_B = 0.488$ degeneracy. The error is somewhat relieved by partial inclusion of the non-local exact exchange in PBE0, while it is completely corrected by the full non-local exact exchange effectively included in PZ-SIC+\rrscanc. While at dissociation this problem of spurious fractional occupation can be avoided by imposing the condition of integer occupation on the solutions, the same is not possible at finite separations which are more relevant to real materials.  

\begin{figure}
    \centering
    \includegraphics[width=\textwidth]{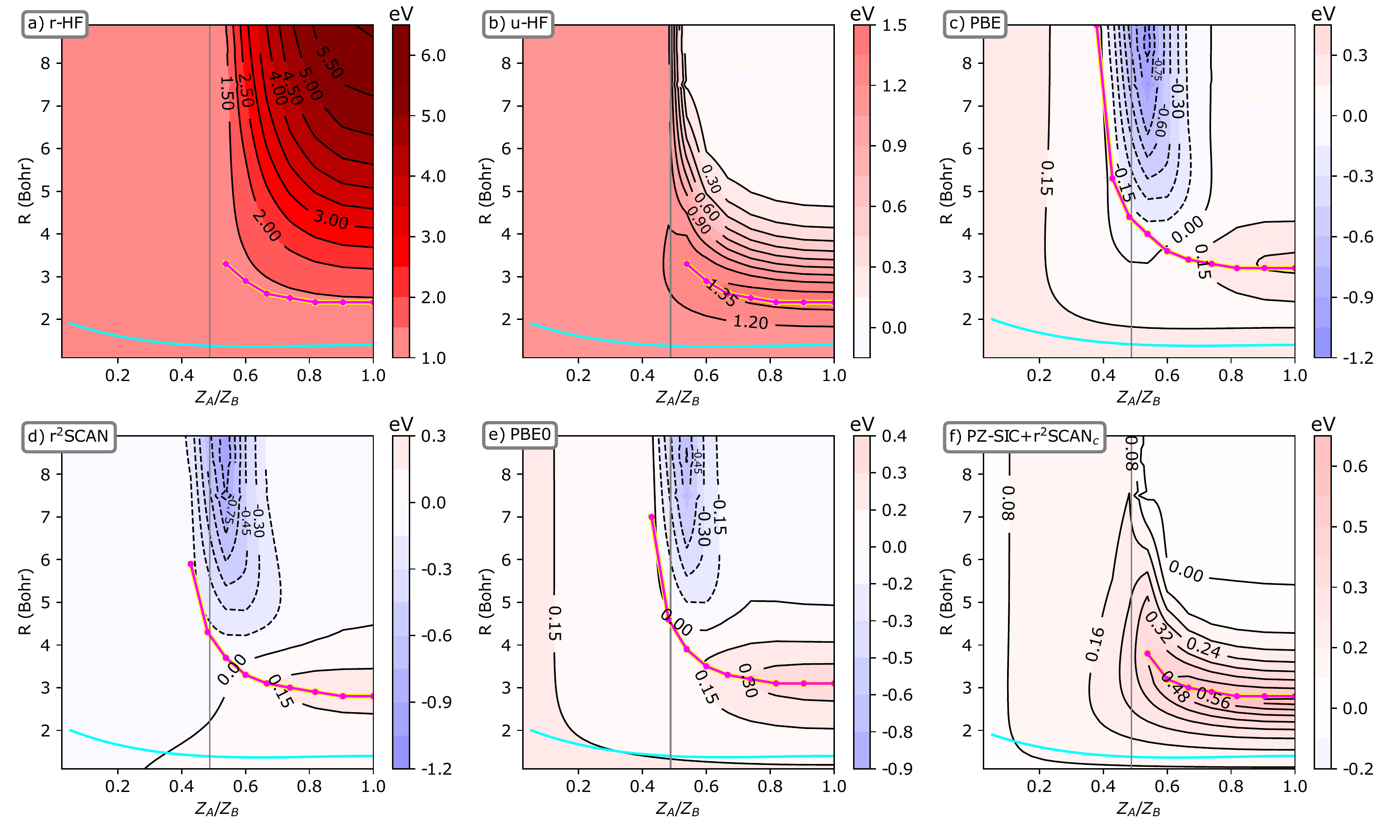}
    
    \caption{\textbf{The errors of different methods in H\cmol{2}{FNC} ground state total energy across the asymmetric nuclear charge ratio for finite bond lengths.} Error is computed from self-consistent calculations relative to basis set exact CCSD (d-aug-cc-pVQZ basis set \cite{Woon1994}) values. Spin symmetry is conserved (r-) for a), and broken (u-) for b-f). The equilibrium bonding distance as predicted by each approximate method are highlighted with a cyan line, and the $Z_A/Z_B = 0.488$ degeneracy point at infinite separation is shown with a gray line. The Coulson--Fisher point at which the symmetry breaking and conserving solutions split for each approximation is shown by a magenta line by sampling points along the surface until no splitting could be found. A common color scale of +6 eV (red) to -6 eV (blue) is used for all plots.}
    \label{fig:energy_landscapes}
\end{figure}

We now turn to H\cmol{2}{FNC} at finite separations. Figure \ref{fig:energy_landscapes} shows the landscapes of energy errors relative to exact CCSD from the methods considered as functions of the $Z_A/Z_B$ ratio and the bond length. The equilibrium bond length for each $Z_A/Z_B$ is highlighted as a cyan line. Comparing r-HF (a) and u-HF (b) reveals how strong correlation is reduced when either bond length or $Z_A/Z_B$ decreases. Both r-HF and u-HF perform similarly once both electrons localize into either the bonding region when the two nuclei are around the equilibrium and $Z_A/Z_B > 0.488$, or onto $Z_B$ when $Z_A/Z_B< 0.488$. Under these conditions the strong correlation caused by spin symmetry is significantly reduced, and most of the error is the result of missing short-range dynamic correlation. 

It is well known from studies of H\mol{2} that the symmetry conserving and symmetry breaking solutions separate at the Coulson--Fischer point \cite{Coulson1949}, around 2.4 Bohr for H\mol{2}. The position of this separation is strongly affected by the nuclear charge asymmetry, occurring at longer bond lengths as $Z_A/Z_B$ decreases. Below the $Z_A/Z_B = 0.488$ degeneracy, the Coulson--Fischer point disappears as the u-HF and r-HF solutions mostly coincide, as shown in Figure \ref{fig:energy_landscapes} (a-b). We expect that the Coulson--Fischer point stretches to the infinity bond length at the $Z_A/Z_B=0.488$ degeneracy if calculations with denser $Z_A/Z_B$ and bond length grids can be converged. It is interesting to note that the maximum error for u-HF tracks the Coulson--Fischer point with $Z_A/Z_B$. 

Figures \ref{fig:energy_landscapes} (c)-(f) show that in general the DFT approximations are a significant improvement over HF as a result of their ability to capture the short-range dynamic correlation. Two important error patterns develop however. One is the underestimation of total energy at the region of long bond lengths ($>$ 5 Bohr) centered around the $Z_A/Z_B = 0.488$ where the SO-DO degeneracy occurs at infinite separation. This error appears to have the same origin as that at infinite separation. PZ-SIC+\rrscanc removes the self-interaction error and gives only negligible errors at long bond lengths for the whole range of $Z_A/Z_B$. Note that unlike at infinite separation it is not possible to enforce integer occupations onto nuclei for approximate calculations, as the electrons can delocalize across both centers freely in the exact solution.

The other error pattern tracks the Coulson--Fischer point with $Z_A/Z_B$ and is similar in nature to the u-HF error maximum. The error is roughly proportional to the percentage of the non-local exact exchange included in the DFT approximations, with the maximum errors of 0.65 eV for PZ-SIC+\rrscanc (100$\%$), 0.4 eV for PBE0 (25$\%$), 0.35 eV for PBE (0$\%$), and 0.3 eV for \rrscan (0$\%$). The preference for a smaller fraction of exact exchange in this error pattern can be explained from the error cancellation between the exchange and correlation approximations \cite{Sun2015}, which is required for good performance for normal materials. The error pattern disappears when $Z_A/Z_B < 0.488$ as both electrons localize on the same nucleus. This indicates the error has a multi-center origin that is non-local and driven by the emerging strong correlation accompanied by the Coulson--Fischer point. Because the 100$\%$ nonlocal exact exchange in PZ-SIC+\rrscanc can not take advantage of the error cancellation with the semilocal \rrscanc correlation for modeling the emerging non-local strong correlation, the error develops more strongly for PZ-SIC+\rrscanc than for the other DFT approximations in the longer bond length domain when $Z_A/Z_B > 0.488$. Interestingly, the self-interaction error that removes the discontinuity in the ground state energy around $Z_A/Z_B \approx 0.488$ at infinite separation results in electron density leaking onto the less charged $Z_A$, allowing the Coulson--Fischer point to persist when $Z_A/Z_B < 0.488$. Given the good performance of PZ-SIC+\rrscanc at infinite separation, this highlights the challenging problem of delivering accuracy for both regions dictated by self-interaction errors and multi-center non-local strong correlation. We therefore expect that H\cmol{2}{FNC} can be a powerful tool for developing the non-local density functionals that have been the focus of much recent DFT development \cite{Klupfel2012, Pederson2014, Vydrov2006, Cohen2008b, Maier2018, Becke2003, Zope2019a}.

It is well accepted that DFT with sophisticated exchange-correlation approximations have better accuracy than HF, and that accuracy generally improved when climbing up the the Perdew--Schmidt hierarchy, e.g., from PBE, to \rrscan, and to PBE0. This is consistent with the observation in Figure \ref{fig:energy_landscapes} that general performance is improved from HF, to PBE, to \rrscan, and to PBE0, shown by smaller error scales and overall smaller regions of error. Similarly, PZ-SIC has been shown as an effective correction to DFT approximations for correlated materials due to the removal of self-interaction errors \cite{Strange1999, Szotek1993}. Correcting DFT with PZ-SIC deteriorates accuracy for normal materials however, an effect which has been called ``the paradox for PZ-SIC'' \cite{Perdew2015}. This agrees with the increased error found around the Coulson--Fischer point for PZ-SIC+\rrscanc in Figure \ref{fig:energy_landscapes} (f). We shall use the performance of DFT approximations for 3$d$ valence transition metal monoxides as an illustration of such connections.

Table \ref{tab:TMO} shows the predicted electronic band gaps and magnetic moments for four typical 3$d$ binary oxide antiferromagnetic (AFM) insulators (MnO, FeO, and CoO, NiO), materials which led to the initial understanding of strongly correlated Mott insulators through on-site correlation localizing electrons into $d$ bands \cite{Mott1974}. Density functional methods have typically struggled with such materials, suffering from the self-interaction error that leads to a spurious charge delocalization between the metal and oxygen ions \cite{Zhang2020b}. The delocalization error enhances the overlap between the $d$ orbitals of metal ions and $p$ orbitals of oxygen ions, and thus destabilizes the magnetic moments of metal ions, which results in too small band gaps. The same tendency for charge delocalization is also observed in Figure \ref{fig:energy_landscapes} for H\cmol{2}{FNC} around the $Z_A/Z_B = 0.488$ degeneracy. Comparing the predictions in Table \ref{tab:TMO} with the range of errors observed in Figure \ref{fig:energy_landscapes}, we see the large region of delocalization error for PBE are reflected in underestimated magnetic moments and qualitatively incorrect band gaps. The region of delocalization error is smaller for \rrscan and correspondingly the material predictions are improved, with all materials correctly insulating though significant underestimation of band gaps remains. The partial self-interaction error correction from the non-local exchange admixture in PBE0 further reduces delocalization error and generally improves magnetic moments and band gaps. Early work with PZ-SIC-corrected LSDA had predictions similar to PBE0 \cite{Szotek1993}.

\begin{table}
    \caption{\textbf{Comparison of theoretically predicted band gaps and local magnetic moments for four 3$\mathbf{d}$ transition metal monoxides.} Experimental ionic positions and lattice constants are used, with experimental reference data from Ref.~\cite{Zhang2020b} and references therein. } 
    \label{tab:TMO}
    \centering
    \begin{tabular}{c|cccc|cccc}\hline\hline
         &\multicolumn{4}{c|}{Band gap (eV)} & \multicolumn{4}{c}{Magnetic moment ($\mu_B$)} \\
         \hline
         Structure & MnO & FeO & CoO & NiO & MnO  & FeO & CoO & NiO \\ \hline
         PBE & 0.91 & 0.00 & 0.00 & 1.03 & 4.33 & 3.46 & 2.43 & 1.37 \\
         \rrscan & 1.69 & 0.59 & 0.89 & 2.50 & 4.45  & 3.56 & 2.58 & 1.59\\
         PBE0 & 3.66  & 3.06 & 4.30 & 5.25 & 4.53  & 3.66 & 2.68 & 1.68\\
         Expt.~\cite{Zhang2020b} & 3.5 & 2.4 & 2.8  & 4.0-4.3 & 4.58 & 4.0 & 3.8 & 1.9\\
        \hline\hline
    \end{tabular}
    \\
\end{table}

Finally, we would like to highlight how the fractional nuclear charges open a new door for understanding electron behavior and electronic structure theory. The 2-electron H\cmol{2}{FNC} system studied here informs single orbital performance and can easily be extended to 3 or more electrons to capture multi-orbital physics. The neutral 3-electron fractional nuclear charge diatomic can be considered as a direct analogue of the ``two center, three electron'' bonding in homo-nuclear X\cmol{2}{+} diatomic cations \cite{Zhang1998}. The study of the charge transfer and strong correlation competition in multi-orbital systems can be conducted by allowing fractional nuclear charges for, e.g., ``Cr\cmol{2}{FNC}''. Additionally, one could increase the number of nuclear centers present, though the spatial arrangements become less simple. A 1D chain presents $Z_A$-H\mol{n}-$Z_B$ arrangements where $Z_A/Z_B$ is tuned such that a small perturbation drives transitions between single and double occupation ground states, reflecting charge transfer across intermediate orbitals. Alternatively, a chain of fractionally-charged nuclear pairs can be extended to infinity, $(Z_A-Z_B)_\infty$, or arranged into 2 and 3 dimensional lattices, analogous to Hubbard models. Naturally, obtaining accurate reference solutions becomes more expensive as the number of electrons increases. 

Despite its simplicity the 2-electron H\cmol{2}{FNC} paradigm system offers a rich window into the competing strong correlation and charge transfer physics that drive the exotic properties of many complex materials. We have shown how calculations of H\cmol{2}{FNC} at finite and infinite nuclear separations highlight important deficiencies in DFT approximations. This identified two major error sources originating from self-interaction error and the multi-center non-local correlation accompanying the Coulson--Fischer point where the spin symmetry breaking and conserving solutions meet. None of the DFT approximations considered could remove both error sources, even when spin symmetry breaking was applied. These errors were connected to accuracy for transition metal monoxides, showing their importance for predicting properties of real materials. The H\cmol{2}{FNC} presented can be easily extended to more complex multi-orbital systems, offering a clear and practical sandbox for one of the largest problems remaining in the physical sciences.

\section{Acknowledgments}

J.W.F., R.Z., and J.S. acknowledge the support of the U.S. DOE, Office of Science, Basic Energy Sciences Grant No. DE-SC0019350 (core research). We thank John Perdew and Lin Hou for their comments.

\section{Author Contributions}

J.W.F. and R.Z. performed calculations and analyzed data. J.W.F. and J.S. designed and led the investigations, designed the computational approaches, analyzed results, and wrote the manuscript. J.S. provided computational resources. All authors contributed to editing the manuscript.

\section{Data Availability Statement}

Data for Figures 1 and 2 is available from the authors by request.

\section{Methods}

\subsection{Fractional Nuclear Charges}

Fractional nuclear charges are implemented under the Born--Oppenheimer approximation by assigning desired $Z_i \in \mathbb{R}^+$ to each nucleus and evaluating the nuclear-electron attraction and nuclear repulsion integrals in the standard way. This modification is trivial for most existing electronic structure codes and is available in the standard \textsc{Turbomole} release used for this work \cite{TURBOMOLE, Turbomole2020}.

\subsection{Figure \ref{fig:frac_atoms}}

Total energies calculated using \textsc{Turbomole} V7.4 as the sum of two independent atomic calculations with fractional nuclear charges. The d-aug-cc-pV5Z hydrogen basis functions \cite{Woon1994} were used for all atomic calculations. Fractional electron occupation was determined numerically by adjusting the occupation fraction on each atomic fragment (fixed such that the total system contains two electrons) to minimize self-consistent total energy.

\subsection{Figure \ref{fig:energy_landscapes}}

Total energies calculated for fractional nuclei at finite separation compared to CCSD references, all calculated using \textsc{Turbomole} V7.4. The d-aug-cc-pVQZ hydrogen basis functions \cite{Woon1994} were used for all calculations, no basis set superposition error (BSSE) corrections were applied. Coulson--Fisher points were evaluated at regular steps in $Z_A/Z_B$ by numerically searching for the minimum bond length $R$ ($\pm0.05$ Bohr) where the spin restricted and spin unrestricted total energies differed by $> 10^{-4}$ eV. No point was recorded if no separation was found $< 9$ Bohr.

\subsection{Table \ref{tab:TMO}}
All materials are in the $G$-type AFM phase. Calculations use the pseudopotential projector-augmented wave method~\cite{Kresse1999} as implemented in the Vienna {\it ab initio} simulation package (VASP)~\cite{Kresse1993,Kresse1996}. A high-energy cutoff of 500 eV was used to truncate the plane-wave basis set.

%

\end{document}